\documentclass[conference]{IEEEtran}

\ifCLASSINFOpdf
\else
   \usepackage[dvips]{graphicx}
\fi
\usepackage{url}
\usepackage{graphicx}
\usepackage{color}
\usepackage{cite}
\usepackage{amsmath}
\usepackage{amssymb}
\usepackage{subfigure}
\usepackage{algorithm}
\usepackage{algorithmic}
\usepackage{enumitem}
\IEEEoverridecommandlockouts
\begin{document}

\title{Jointly Sparse Signal Recovery via Deep Auto-Encoder and Parallel Coordinate Descent Unrolling}

\author{
\IEEEauthorblockN{Shuaichao~Li, Wanqing~Zhang, Ying~Cui} 
\IEEEauthorblockA{Dept. of EE, Shanghai Jiao Tong University, China}
\thanks{S.~Li and W.~Zhang contributed equally. This work was supported in part by the National Key R\&D Program of China under Grant 2018YFB1801102.}
}

\maketitle

\begin{abstract}
In this paper, utilizing techniques in compressed sensing, parallel optimization and deep learning, we propose a model-driven approach to jointly design the common measurement matrix and GROUP LASSO-based jointly sparse signal recovery method for complex sparse signals, based on the standard auto-encoder structure for real numbers. The encoder achieves noisy linear compression for jointly sparse signals, with a common measurement matrix. The GROUP LASSO-based decoder realizes jointly sparse signal recovery based on an iterative parallel-coordinate descent (PCD) algorithm which is proposed to solve GROUP LASSO in a parallel manner. In particular, the decoder consists of an approximation part which unfolds (several iterations of) the proposed iterative algorithm to obtain an approximate solution of GROUP LASSO and a correction part which reduces the difference between the approximate solution and the actual jointly sparse signals. The proposed model-driven approach achieves higher recovery accuracy with less computation time than the classic GROUP LASSO method, and the gain significantly increases in the presence of extra structures in sparse patterns. The common measurement matrix obtained by the proposed model-driven approach is also suitable for the classic GROUP LASSO method.  We consider an application example, i.e., channel estimation in Multiple-Input Multiple-Output (MIMO)-based grant-free random access which is proposed to support massive machine-type communications (mMTC) for Internet of Things (IoT). By numerical results, we demonstrate the substantial gains of the proposed model-driven approach over GROUP LASSO and AMP when the number of jointly sparse signals is not very large.
\end{abstract}
\begin{IEEEkeywords}
Compressed sensing, jointly sparse signal recovery, grant-free random access, channel estimation, GROUP LASSO, deep learning, auto-encoder, parallel optimization.
\end{IEEEkeywords}
\section{Introduction}
Jointly sparse signal recovery in Multiple Measurement Vector (MMV) models refers to the estimation of $M$ jointly sparse $N$-dimensional signals from $L$ ($\ll$$N$) limited noisy linear measurements based on a common measurement matrix. When $M=1$, jointly sparse signal recovery reduces to sparse signal recovery in Single Measurement Vector (SMV) models. The jointly sparse signal recovery problem arises in many applications in communications and signal processing. Two main challenges exist in jointly sparse signal recovery. One is to design a common measurement matrix which maximally retains the information on jointly sparse signals when reducing signal dimension. The other is to recover the jointly sparse signals with high recovery accuracy and short computation time.

Existing works on the jointly sparse signal recovery in MMV models investigate the estimation of jointly sparse signals for a given common measurement matrix \cite{qin2013efficient,8323218, 8264818,senel2018grant}. Pure jointly sparse signal recovery design includes optimization-based methods such as GROUP LASSO \cite{qin2013efficient} and iterative thresholding methods such as AMP \cite{8323218,8264818,senel2018grant}. Note that GROUP LASSO reduces to LASSO when $M=1$. A closely related topic is to recover the common support of jointly sparse signals for a given common measurement matrix \cite{koochakzadeh2018fundamental,8437359,obozinski2011support,liu2018sparse}. Specially, pure jointly sparse support recovery design includes exhaustive methods \cite{koochakzadeh2018fundamental}, optimization-based methods such as Maximum Likelihood (ML) estimation \cite{8437359} and GROUP LASSO \cite{obozinski2011support}, and iterative thresholding methods such as AMP \cite{liu2018sparse}. On one hand, given jointly sparse signals, one can set a threshold on them to obtain the common support. On the other hand, given a common support, the estimation of the jointly sparse signals can be done by the classic Minimum Mean Squared Error (MMSE) method \cite{8437359}. It is worth noting that none of \cite{qin2013efficient,8323218,8264818,senel2018grant,koochakzadeh2018fundamental,8437359,obozinski2011support,liu2018sparse} considers design of the common measurement matrix, or exploits characteristics of sparse patterns for improving recovery accuracy. Hence, the proposed methods in \cite{qin2013efficient,8323218,8264818,senel2018grant,koochakzadeh2018fundamental,8437359,obozinski2011support,liu2018sparse} may not achieve desirable recovery performance.

It has been shown that joint design of signal compression and recovery methods for real signals \cite{sun2016deep,wu2019learning,nguyen2017deep,8262812,shi2017deep,adler2017block} or complex
signals \cite{8322184} using auto-encoder in deep learning can significantly improve recovery performance. However, \cite{sun2016deep,wu2019learning,nguyen2017deep,8262812,shi2017deep,adler2017block,8322184} are not for the MMV models, and their extensions to MMV models are highly nontrivial. In addition, note that neither
the neural network for complex signals in \cite{8322184} nor direct extensions of the neural networks for real signals in \cite{sun2016deep,wu2019learning,nguyen2017deep,8262812,shi2017deep,adler2017block} to complex signals can achieve linear compression for complex signals. The authors in \cite{wu2019learning} approximately unfold the projected sub-gradient algorithm for LASSO without considering the effect of noise. As the adopted projected sub-gradient algorithm involves matrix inverse, which cannot be implemented exactly using a neural network, the resulting recovery performance will degrade. 

In our previous work \cite{8861085}, an auto-encoder-based approach is proposed to jointly design the measurement matrix (achieving linear compression) and support recovery method for sparse complex signals in SMV models. Note that directly extending the decoder for the SMV models in \cite{8861085} to MMV models will not provide promising recovery performance. The MMSE method involves matrix inversion, and cannot be implemented precisely using a neural network. Thus, it is not feasible to obtain an auto-encoder-based approach for jointly sparse signal recovery in MMV models, by extending the proposed approach for sparse support recovery in \cite{8861085} and using the MMSE method. In addition, unlike the space for support, the space for signals is mostly infinite. Thus, approximating jointly sparse signal recovery based on only training samples probably requires huge computing resources and extensive computation time, and may yield unsatisfactory recovery accuracy. Thus, how to jointly design the common measurement matrix and jointly sparse signal method for complex signals in MMV models remains open.

In this paper, utilizing techniques in compressed sensing, parallel optimization, and deep learning, we propose a model-driven approach to jointly design the common measurement matrix and jointly sparse signal recovery method for complex signals, based on the standard auto-encoder structure for real numbers. It is worth noting that the proposed model-driven approach is applicable for real signals, which are special cases of complex signals. Specifically, we build an encoder to mimic the noisy linear measurement process for jointly sparse signals with a common measurement matrix. We build a GROUP LASSO-based decoder which consists of an approximation part and a correction part, to realize jointly sparse signal recovery based on an iterative algorithm that is proposed to solve GROUP LASSO in a parallel manner via the parallel-coordinate descent (PCD) method. Specifically, the approximation part unfolds (several iterations of) the proposed PCD algorithm to obtain an approximate solution of GROUP LASSO, and the correction part reduces the difference between the approximate solution and the actual jointly sparse signals. The proposed model-driven approach can effectively utilize the properties of sparsity patterns, and is especially useful when it is hard to analytically model the underlying structures of sparsity patterns. In addition, the proposed model-driven approach achieves higher recovery accuracy with short computation time than the classic GROUP LASSO method \cite{qin2013efficient}. We consider an important application example, i.e., channel estimation in MIMO-based grant-free random access, which is proposed to support massive machine-type communications (mMTC) for Internet of Things (IoT) \cite{8323218,8264818,senel2018grant,8437359,liu2018sparse}, and apply our proposed model-driven approach therein. By extensive numerical results, we demonstrate the substantial gains of the proposed model-driven approach over GROUP LASSO \cite{qin2013efficient} and AMP \cite{8323218} in terms of signal recovery accuracy, especially when the number of jointly sparse signals is not very large.

\textbf{Notation:} We represent vectors by boldface small letters (e.g., $\mathbf{x}$), matrices by boldface capital letters (e.g., $\mathbf{X}$), scalar constants by non-boldface letters (e.g., $x$ or $X$) and sets by calligraphic letters (e.g., $\mathcal{X}$). The notation $X(i,j)$ denotes the $(i,j)$-th element of matrix $\mathbf{X}$, $\mathbf{X}_{i,:}$ represents the $i$-th row of matrix $\mathbf{X}$, $\mathbf{X}_{:,i}$ represents the $i$-th column of matrix $\mathbf{X}$, and $x(i)$ represents the $i$-th element of vector $\mathbf{x}$. Superscript $^H$ and superscript $^T$ denote transpose conjugate and transpose, respectively. The notation $\||\mathbf{X}\||_F$ represents the Frobenius norm of matrix $\mathbf{X}$, and $\Re(\cdot)$ and $\Im(\cdot)$ represent the real part and imaginary part, respectively. $\mathbf{0}_{m\times n}$ and $\mathbf{I}_{n\times n}$ represent the $m\times n$ zero matrix and the $n\times n$ identity matrix, respectively. The complex field and real field are denoted by $\mathbb{C}$ and $\mathbb{R}$, respectively.

\section{Jointly Sparse Signal Recovery and Application}
\label{sec:format}

The support of a sparse $N$-dimensional complex signal $\mathbf{x} \triangleq (x(n))_{n \in \mathcal{N}}\in \mathbb{C}^N$ is defined as the set of locations of non-zero elements of $\mathbf{x}$, and is denoted by ${\rm supp}(\mathbf{x}) \triangleq \{n\in\mathcal{N}|x(n) \neq 0\}$, where $\mathcal{N} \triangleq \{1,\cdots,N\}$. If the number of non-zero elements of $\mathbf{x}$ is much smaller than its total number of elements, i.e., $|{\rm supp}(\mathbf{x})| \ll N $, $\mathbf{x}$ is sparse. Consider a set of $M$ jointly sparse signals $\mathbf{x}_m\in \mathbb{C}^N, m \in \mathcal{M} \triangleq \{1,\cdots,M\}$, sharing a common support $\mathcal{S} \triangleq {\rm supp}(\mathbf{x}_m), m \in \mathcal{M}$. For all $m\in \mathcal{M}$, consider $L \ll N$ noisy linear measurements $\mathbf{y}_m\in \mathbb{C}^L$ of $\mathbf{x}_m$, i.e., $\mathbf{y}_m = \mathbf{A}\mathbf{x}_m+\mathbf{z}_m$, where $\mathbf{A}\in \mathbb{C}^{L\times N}$ is the common measurement matrix, and $\mathbf{z}_m\sim\mathcal{CN}(\mathbf{0}_{L\times 1},\sigma^2\mathbf{I}_{L\times L})$ is the additive white Gaussian noise. More compactly, define $\mathbf{X} \in \mathbb{C}^{N\times M}$ with $\mathbf{X}_{:,m} \triangleq \mathbf{x}_m,\; m\in \mathcal{M}$, $\mathbf{Y}\in \mathbb{C}^{L\times M}$ with $\mathbf{Y}_{:,m} \triangleq \mathbf{y}_m,\; m\in \mathcal{M}$ and $\mathbf{Z}\in \mathbb{C}^{L\times M}$ with $\mathbf{Z}_{:,m} \triangleq \mathbf{z}_m,\; m\in \mathcal{M}$. Then, we have:
\begin{align}
\mathbf{Y} =\mathbf{A}\mathbf{X}+\mathbf{Z}
\end{align}
The jointly sparse signal recovery problem in MMV models aims to estimate the $M$ jointly sparse signals $\mathbf{x}_m,m \in \mathcal{M}$ (i.e., $\mathbf{X}$) from $M$ noisy linear measurement vectors $\mathbf{y}_m,m \in \mathcal{M}$ (i.e., $\mathbf{Y}$), obtained through common measurement matrix $\mathbf{A}$.

\begin{figure*}[ht]
\begin{center}
 \resizebox{16cm}{!}{\includegraphics{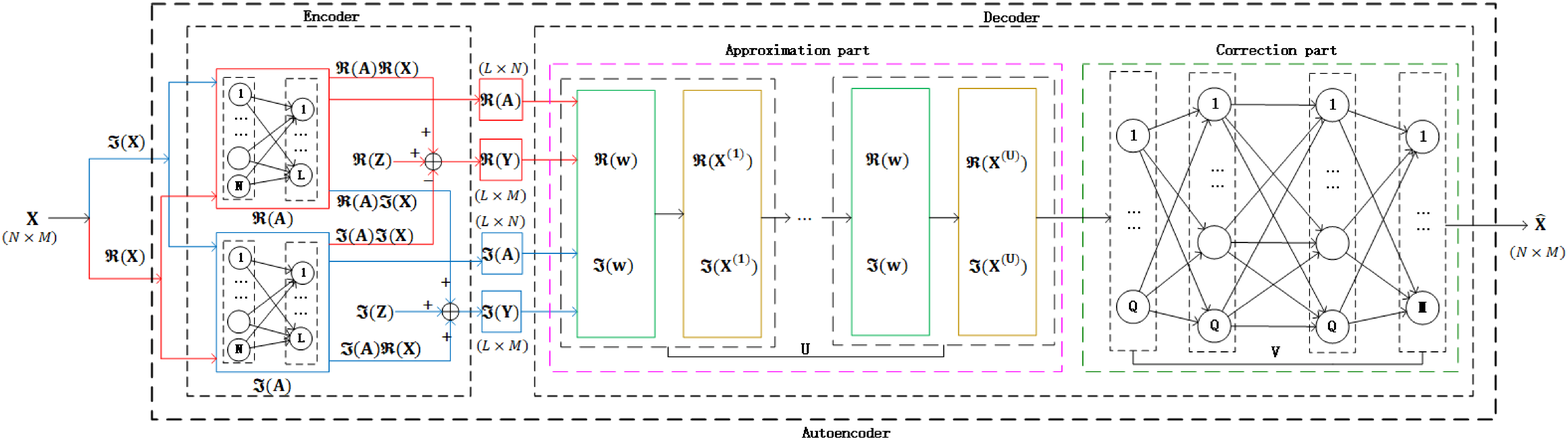}}
  \end{center}
     \caption{Proposed model-driven approach.}
\label{networkmodel1}
\end{figure*}

As an important application example, we consider channel estimation in MIMO-based grant-free random access. Consider a single cell with one $M$-antenna base station (BS) and $N$ single-antenna devices. Let $\alpha_n \in \{0,1\}$ denote the active state of device $n$, where $\alpha_n=1$ means that device $n\in\mathcal{N}$ accesses the channel (i.e., is active), and $\alpha_n=0$ otherwise. For all $m \in \mathcal{M}$, let $h_{m}(n)\in \mathbb{C}$ denote the state of the wireless channel between the $m$-th antenna at the BS and device $n$, which is a complex number, and view $\alpha_nh_{m}(n)$ as $x_m(n)$. That is, $x_{m}(n)$ represents the corresponding channel state if it is not zero. As device activity patterns for IoT traffic are typically sporadic, $\mathbf{x}_{m}\in \mathbb{C}^N,m \in \mathcal{M}$ are jointly sparse with common support $\mathcal{S}=\{n \in \mathcal{N}|\alpha_n=1\}$. In grant-free random access, each device $n$ has a unique pilot sequence $\mathbf{a}_n \in \mathbb{C}^L$, with $L \ll N$. View $\mathbf{A} \in \mathbb{C}^{L\times N}$ with $\mathbf{A}_{:,n} = \mathbf{a}_n, n\in \mathcal{N} $ as the pilot matrix, which is known at the BS. In the pilot transmission phase, active devices synchronously send their pilot sequences to the BS. Then, $\mathbf{Y}$ in (1) represents the received signal at the BS. The BS conducts channel estimation by estimating $\mathbf{x}_{m}\in \mathbb{C}^N,m \in \mathcal{M}$ \big(i.e., $\mathbf{X}$\big) from $\mathbf{Y}$, given knowledge of $\mathbf{A}$, which obviously corresponds to jointly sparse signal recovery in MMV models.

\section{Proposed Approach}
\label{approach}
In this section, a model-driven approach is proposed to jointly design the common measurement matrix and jointly sparse signal recovery method for sparse complex signals in MMV models. As illustrated in Fig.~\ref{networkmodel1}, the proposed approach is based on auto-encoder for real numbers in deep learning.


\subsection{Encoder}
The encoder mimics the noisy linear measurement process for jointly sparse signals with a common measurement matrix in (1). To mimic (1) using standard auto-encoder for real numbers in deep learning, we equivalently express (1) as:
\begin{align}
\Re(\mathbf{Y})=\Re(\mathbf{A})\Re(\mathbf{X})-\Im(\mathbf{A})\Im(\mathbf{X})+\Re(\mathbf{Z})\\
\Im(\mathbf{Y})=\Im(\mathbf{A})\Re(\mathbf{X})+\Re(\mathbf{A})\Im(\mathbf{X})+\Im(\mathbf{Z})
\end{align}
Two fully-connected neural networks, each with two layers, are built to implement multiplications with matrices $\Re(\mathbf{A})\in \mathbb{R}^{L\times N}$ and $\Im(\mathbf{A})\in \mathbb{R}^{L\times N}$, respectively. For each neural network, there are $N$ neurons and $L$ neurons in the input layer and the output layer, respectively; the weight of the connection from the $n$-th neuron in the input layer to the $l$-th neuron in the output layer corresponds to the $(l,n)$-th element of the corresponding matrix; and no activation functions are used in the output layer. The elements of $\Re(\mathbf{Z})\in \mathbb{R}^{L\times M}$ and $\Im(\mathbf{Z})\in \mathbb{R}^{L\times M}$ are generated independently according to $\mathcal{N}(0,\frac{\sigma^2}{2})$. As shown in Fig.~\ref{networkmodel1}, when $\Re(\mathbf{X})\in \mathbb{R}^{N\times M}$ and $\Im(\mathbf{X})\in \mathbb{R}^{N\times M}$ are input to the encoder, $\Im(\mathbf{Y})\in \mathbb{R}^{L\times M}$ and $\Re(\mathbf{Y})\in \mathbb{R}^{L\times M}$ can be easily obtained. Note that the encoder can be viewed as an extension of the one in our previous work \cite{8861085}.
\subsection{Decoder}
The decoder minics the jointly sparse signal recovery process based on GROUP LASSO. Note that jointly sparse signal recovery is in general more challenging than jointly sparse support recovery, as the space of sparse signals is much larger than the space of support, and is even infinite in some scenarios. Approximating the jointly sparse signal recovery based only on training samples probably requires huge computing resource and extensive computation time, and may yield unsatisfactory recovery accuracy. Thus, we propose a GROUP LASSO-based decoder which consists of two parts. The first part is to obtain an approximate solution of GROUP LASSO, and is called the approximation part. The second part is to reduce the difference between the
approximate solution and the actual jointly sparse signals, and is referred to as the correction part.


First, we introduce GROUP LASSO, which is a natural optimization-theoretic formulation of the jointly sparse signal recovery problem in MMV models \cite{qin2013efficient}. The formulation of GROUP LASSO is as below:
\vspace{-0.3cm}
\begin{align}
\label{gl}
\min_\mathbf{X}\;(1/2)\||\mathbf{A}\mathbf{X}-\mathbf{Y}\||_F^2+\lambda\sum_{i=1}^{N}\|\mathbf{X}_{i,:}\|_2
\end{align}
where $\lambda \geq 0$ is a tuning parameter. Let $\mathbf{X}^*$ denote an optimal solution of this problem. Note that when $M=1$, the formulation in (\ref{gl}) reduces to LASSO.

The optimization problem in (\ref{gl}) is convex and can be solved effectively using an iterative algorithm based on the block-coordinate descent method, referred to as BCD-MMV \cite{qin2013efficient}. Specially, each row of $\mathbf{X}$ is treated as one block, and in each iteration, $N$ blocks are updated one after another in a sequential manner.
Define the threshold function as:
\vspace{-0.25cm}
\begin{align}
\label{th}
f(x,\eta)\triangleq\begin{cases}
x-\lambda\eta, &{x > \lambda\eta} \\
0, &{|x| \leq \lambda\eta} \\
x+\lambda\eta, &{x < -\lambda\eta}
\end{cases}
\end{align}
For completeness, we present the details of BCD-MMV in Algorithm 1. It has been shown in \cite{7547360} that $\mathbf{X}^{(k)}\to \mathbf{X}^*$, as $k\to \infty$.
\begin{algorithm}[t]
    \caption{BCD-MMV \cite{qin2013efficient}}
    \begin{algorithmic}[1]
        \STATE Set $\mathbf{X}^{(0)}=\mathbf{0}_{M\times N}$ and $k=1$
        \REPEAT
        \STATE Set $\tilde{\mathbf{X}}=\mathbf{X}^{(k-1)}$.
        \FOR{$i=1,\dots,N$}
        \STATE Compute $\tilde{\mathbf{X}}_{i,:} = \frac{\mathbf{w}}{\|\mathbf{w}\|_2}f(\frac{\|\mathbf{w}\|_2}{\| \mathbf{A}_{:,i}\|_2^2},\frac{1}{\| \mathbf{A}_{:,i}\|_2^2})$, where $f(\cdot,\cdot)$ is given by (\ref{th}), and $\mathbf{w} = \| \mathbf{A}_{:,i}\|_2^2\tilde{\mathbf{X}}_{i,:}-\mathbf{A}_{:,i}^H(\mathbf{A}\tilde{\mathbf{X}}-\mathbf{Y})$.
        \ENDFOR
        \STATE Update $\mathbf{X}^{(k)}=\tilde{\mathbf{X}}$.
        \STATE Set $k=k+1$.
        \UNTIL $k=k_{max}$ or $\mathbf{X}^{(k)}$ satisfies some stopping criterion.
    \end{algorithmic}
\end{algorithm}     

If we directly unfold BCD-MMV in Algorithm 1, $N$ layers are required to implement one iteration. That is, the structure of BCD-MMV does not yield an efficient neural network implementation. To make full use of the parallelizable neural network architecture, we propose an iterative algorithm based on the parallel-coordinate descent method, and is referred to as PCD-MMV. Different from BCD-MMV in Algorithm 1, in each iteration of PCD-MMV, $N$ blocks are updated in parallel. Let $\gamma^{(k)}$ denote a positive diminishing step size satisfying: 
\vspace{-0.22cm}
\begin{align}
\label{size}
&\gamma^{(k)}>0,\ \lim_{k \rightarrow \infty}\gamma^{(k)}=0,\ \sum^\infty_{k=1}\gamma^{(k)}=\infty,\ \sum^\infty_{k=1}(\gamma^{(k)})^2\leq\infty.
\end{align}
The details of PCD-MMV are summarized in Algorithm 2.
\begin{algorithm}[tp]
    \caption{PCD-MMV}
    \begin{algorithmic}[1]
        \STATE Set $\mathbf{X}^{(0)}=\mathbf{0}_{M\times N}$ and $k=1$.
        \REPEAT
        \FOR{$i=1,\dots,N$}
        \STATE Compute $\tilde{\mathbf{X}}_{i,:} = \frac{\mathbf{w}}{\|\mathbf{w}\|_2}f(\frac{\|\mathbf{w}\|_2}{\| \mathbf{A}_{:,i}\|_2^2},\frac{1}{\| \mathbf{A}_{:,i}\|_2^2})$, where $f(\cdot,\cdot)$ is given by (\ref{th}), and $\mathbf{w} = \| \mathbf{A}_{:,i}\|_2^2\mathbf{X}^{(k-1)}_{i,:}-\mathbf{A}_{:,i}^H(\mathbf{A}\mathbf{X}^{(k-1)}-\mathbf{Y})$.
        \ENDFOR
        \STATE Update $\mathbf{X}^{(k)} = \gamma^{(k)}\tilde{\mathbf{X}}+(1-\gamma^{(k)})\mathbf{X}^{(k-1)}$, where $\gamma^{(k)}$ satisfies (\ref{size}).
        \STATE Set $k=k+1$.
        \UNTIL $k=k_{max}$ or $\mathbf{X}^{(k)}$ satisfies some stopping criterion.
    \end{algorithmic}
\end{algorithm}
Following the convergence proof for the parallel successive convex approximation (SCA) algorithm in \cite{7547360}, we can easily show that $\mathbf{X}^{(k)} \rightarrow \mathbf{X}^*$, as $k \rightarrow \infty$.

Next, we introduce the approximate part, which unfolds $U(\geq0)$ iterations of Algorithm 2. The operations for complex numbers in Step 4 of Algorithm 2 can be equivalently expressed as:
\begin{align}
&\Re(\tilde{\mathbf{X}}_{i,:}) = \frac{\Re(\mathbf{w})}{\|\mathbf{w}\|_2}f(\frac{\|\mathbf{w}\|_2}{\| \mathbf{A}_{:,i}\|_2^2},\frac{1}{\| \mathbf{A}_{:,i}\|_2^2})\\
&\Im(\tilde{\mathbf{X}}_{i,:}) = \frac{\Im(\mathbf{w})}{\|\mathbf{w}\|_2}f(\frac{\|\mathbf{w}\|_2}{\| \mathbf{A}_{:,i}\|_2^2},\frac{1}{\| \mathbf{A}_{:,i}\|_2^2})\\
&\Re(\mathbf{w})=\| \mathbf{A}_{:,i}\|_2^2\Re(\mathbf{X}^{(k-1)}_{i,:})-\Re(\mathbf{A}_{:,i}^H(\mathbf{A}\mathbf{X}^{(k-1)}-\mathbf{Y}))\\
&\Im(\mathbf{w})=\| \mathbf{A}_{:,i}\|_2^2\Im(\mathbf{X}^{(k-1)}_{i,:})-\Im(\mathbf{A}_{:,i}^H(\mathbf{A}\mathbf{X}^{(k-1)}-\mathbf{Y}))
\end{align}
where $\Re(\mathbf{A}_{:,i}^H(\mathbf{A}\mathbf{X}^{(k-1)}-\mathbf{Y}))$ and $\Im(\mathbf{A}_{:,i}^H(\mathbf{A}\mathbf{X}^{(k-1)}-\mathbf{Y}))$ can be easily expressed in terms of $\Re(\mathbf{A}), \Re(\mathbf{X}^{(k-1)}), \Re(\mathbf{Y}), \Im(\mathbf{A}), \Im(\mathbf{X}^{(k-1)})$ and $\Im(\mathbf{Y})$. In addition, the operations for complex numbers in Step 6 of Algorithm 2 can be equivalently expressed as:
\begin{align}
&\Re(\mathbf{X}^{(k)}) = \gamma^{(k)}\Re(\tilde{\mathbf{X}})+(1-\gamma^{(k)})\Re(\mathbf{X}^{(k-1)})\\
&\Im(\mathbf{X}^{(k)}) = \gamma^{(k)}\Im(\tilde{\mathbf{X}})+(1-\gamma^{(k)})\Im(\mathbf{X}^{(k-1)})
\end{align}Thus, the operations for complex numbers in Algorithm 2 are readily implemented with operations for real numbers.
As illustrated in Fig.~\ref{networkmodel1}, we build the approximation part with $U$ ($\geq 0$) layers, each realizing one iteration of PCD-MMV in Algorithm 2. We can directly input $\mathbf{Y}$ and $\mathbf{A}$ to the approximation part and obtain $\mathbf{X}^{(U)}$ as the output. Note that $\mathbf{X}^{(U)}$ can be treated as an approximation of $\mathbf{X}^*$ with a negligible approximation error at large $U$.

Then, we introduce the correction part which consists of $V$ ($\geq 1$) fully connected layers. Specially, in each of the first $V-1$ correction layers, rectified linear unit (ReLU) is chosen as the activation function. The last correction layer has $M$ neurons with no activation function for producing output $\hat{\mathbf{X}}$.

Note that $U$ and $V$ are jointly chosen according to the size of the jointly sparse signal recovery problem. In particular, the proposed GROUP LASSO-based decoder reduces to the classic GROUP LASSO method, when $U$ is sufficiently large and $V = 0$ (i.e., only the weights in the encoder are adjustable). In this situation, the proposed model-driven approach is used for designing a common measurement matrix for Algorithm 2. When $U=0$ and $V>0$, the proposed model-driven approach degrades to a purely data-driven one for jointly design of common measurement matrix and jointly sparse signal recovery method. When $U>0$ and $V>0$ are properly chosen according to the size of jointly sparse signal recovery, the proposed GROUP LASSO-based decoder can achieve higher recovery accuracy and shorter computation
time than the classic GROUP LASSO method, which will be illustrated in Section IV.

\subsection{Training Process}
We introduce the training procedure for the auto-encoder. Consider $I$ training samples $\mathbf{X}^{[i]},i=1,\cdots,I$. Let $\hat{\mathbf{X}}^{[i]}$ denote the output of the auto-encoder corresponding to input $\mathbf{X}^{[i]}$. To measure the difference between $\hat{\mathbf{X}}^{[i]}$ and $\mathbf{X}^{[i]}$, we adopt the mean-square error (MSE) loss function:
$${\rm E}((\mathbf{X}^{[i]})_{i=1,\cdots,I},(\hat{\mathbf{X}}^{[i]})_{i=1,\cdots,I})=\frac{1}{NI}\sum_{i=1}^{I}\||\mathbf{X}^{[i]}-\hat{\mathbf{X}}^{[i]}\||_F^2$$
We train the auto-encoder using the ADAM algorithm. After training, we obtain the common measurement matrix particularly for the GROUP LASSO-based decoder, via extracting the weights of the encoder. Furthermore, the GROUP LASSO-based decoder can be directly used for jointly sparse signal recovery. 


\section{Numerical Results}
\label{sim}
In this section, we conduct numerical experiments on channel estimation in MIMO-based grant free random access, which id illustrated in Section II. We choose $\mathbf{h}_m \sim \mathcal{CN}(\mathbf{0}_{N\times 1},\mathbf{I}_{N\times N}), m \in \mathcal{M}$ and $\sigma^2=0.1$. We evaluate the MSE $\frac{1}{NI}\sum_{t=1}^{T}\||\mathbf{X}^{[t]}-\hat{\mathbf{X}}^{[t]}\||_F^2$ and computation time (on the same server) over the same set of $T=10^3$ testing samples. To demonstrate the ability of the proposed model-driven approach for effectively utilizing the properties of sparse patterns, we consider two cases, i.e., the i.i.d. case and the correlated case. In the i.i.d case, $N$ devices randomly access the channel in an i.i.d. manner with access probability $\Pr[\alpha_n=1]=p$, $n \in \mathcal{N}$. In the correlated case, $N$ devices are divided into $G$ groups of the same size $N/G$; the active states of the devices within each group are the same; only one of the $G$ groups is selected to be active uniformly at random; $1/G$ can be viewed as the access probability $p$.

\begin{figure}[t]
\begin{center}
 \subfigure[\scriptsize{MSE versus iteration index $k$.}]
 {\resizebox{4.2cm}{!}{\includegraphics{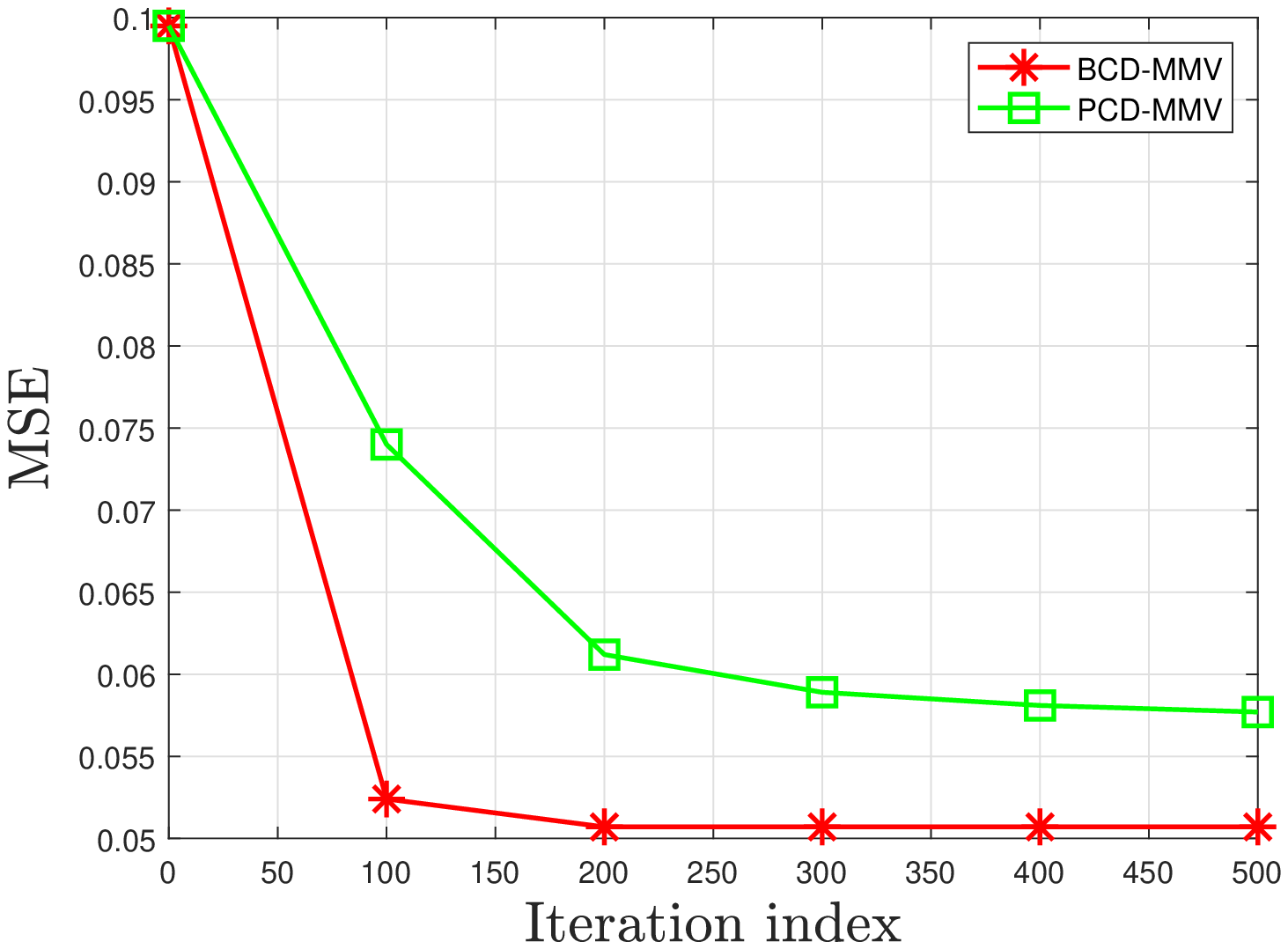}}}
 \subfigure[\scriptsize{Computation time (in seconds) for one iteration.}]
 {\resizebox{4.2cm}{!}{\includegraphics{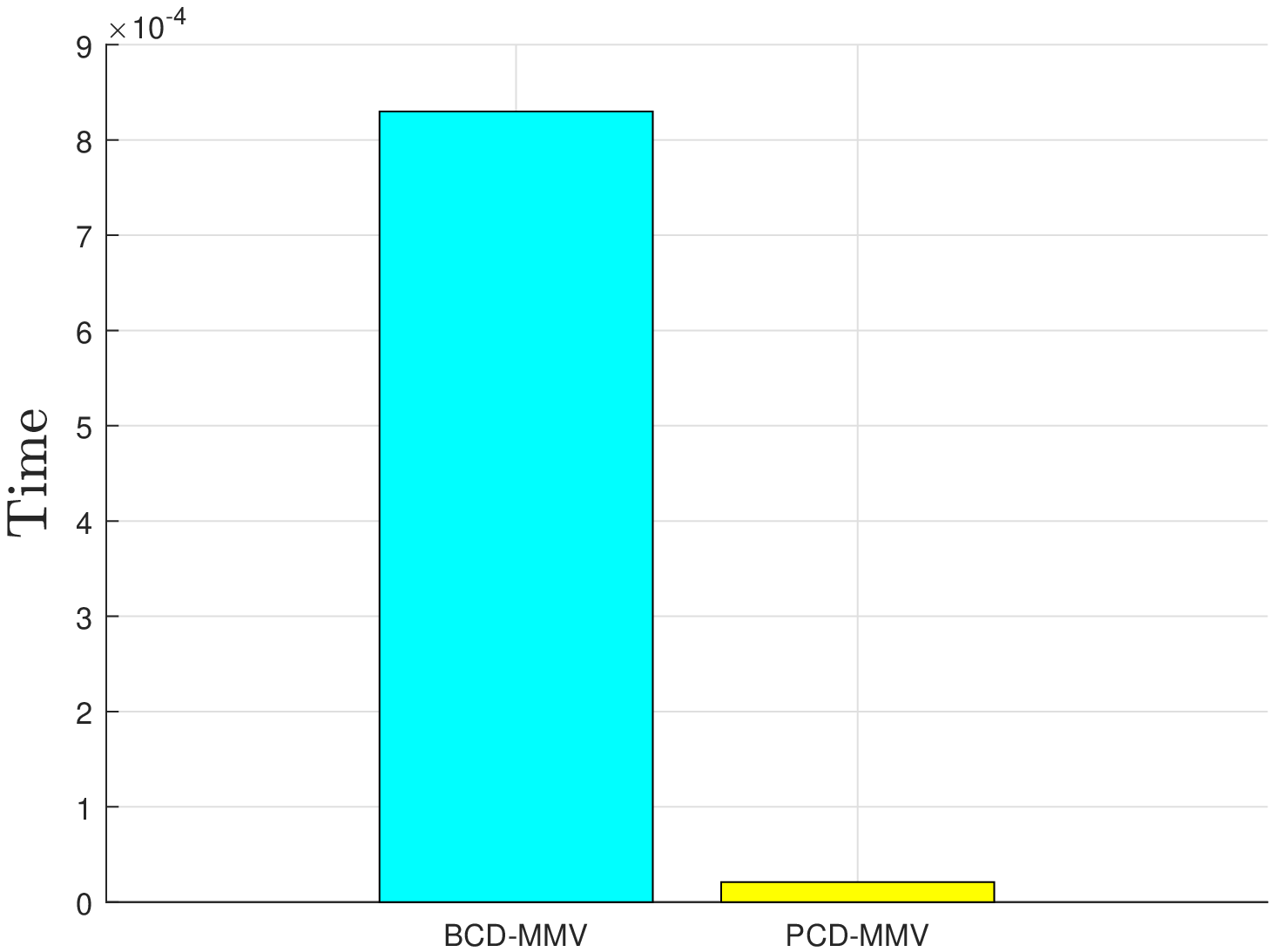}}}
  \end{center}
 \vspace{-0.3cm}
 \caption{\small{MSE versus iteration index ($k$), and computation time at $L=15$, $N=100$, $p=0.1$, $M=4$, $\gamma^{(k)}=\sqrt{k}$.}}
 \label{convergence}
 \vspace{-0.5cm}
\end{figure}

First, we evaluate the convergence and computation time (in seconds) per iteration of BCD-MMV in Algorithm 1 and PCD-MMV in Algorithm 2.\footnote{Note that in Algorithm 2, all $N$ blocks are updated together via matrix operations, for ease of implementation in a neural network.} Here, the entries of the pilot sequences for the $N$ devices are independently generated from $\mathcal{CN}(0,1)$. From Fig.~\ref{convergence}, we can see that BCD-MMV converges much faster than PCD-MMV at the cost of much longer computation time. From Fig.~\ref{convergence}(a), we see that the convergence speed of each algorithm slows down after $k=200$. Thus, in the following, we choose $k_{max}=200$ for both algorithms.

Next, we compare the proposed model-driven approach with four baseline schemes, namely GROUP LASSO (IID), AMP (IID), GROUP LASSO (DL), and AMP (DL). Specifically, GROUP LASSO (IID) and AMP (IID) adopt the same set of pilot sequences for the $N$ devices whose entries are independently generated from $\mathcal{CN}(0,1)$; GROUP LASSO (DL) and AMP (DL) adopt the same set of pilot sequences (corresponding to the weights of the encoder) obtained by the proposed model-driven approach; GROUP LASSO (IID) and GROUP LASSO (DL) conduct channel estimation based on BCD-MMV in Algorithm 1 (with 200 iterations) for GROUP LASSO \cite{qin2013efficient}; and AMP (IID) and AMP (DL) conduct channel estimation using AMP in \cite{8323218} which requires knowledge of the access probability and assumes that devices access the channel in an i.i.d. manner.  For the proposed GROUP LASSO-based decoder, we set $U=200$, $V=3$ and $\gamma^{(k)}=\sqrt{k}$. Those choices for $U$ and $V$ are based on a large number of experiments and the tradeoff between performance and computation time. For fair comparison, we require ${\|\mathbf{a}_n\|}_2=\sqrt{L}$ in training the proposed architecture, as in \cite{8861085}. The training parameters are selected similar to those in \cite{8861085}, and the training details are omitted due to page limitation.


\begin{figure}[t]
\begin{center}
 \subfigure[\scriptsize{MSE versus $L/N$ at $p=0.1$, $M=4$ in the i.i.d case.}]
 {\resizebox{4.2cm}{!}{\includegraphics{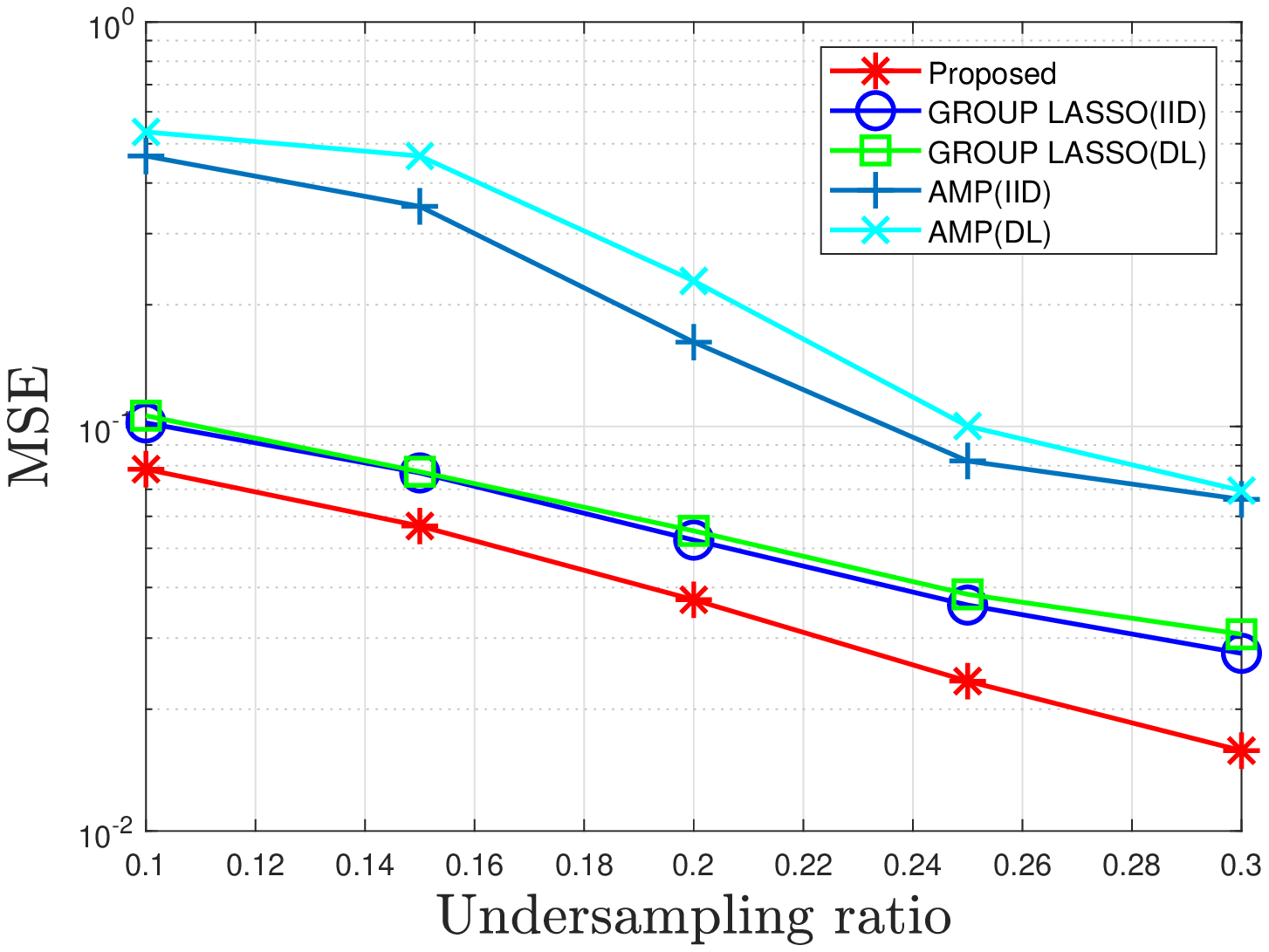}}}
 \subfigure[\scriptsize{MSE versus $p$ at $L/N=0.3$, $M=4$ in the i.i.d. case.}]
 {\resizebox{4.2cm}{!}{\includegraphics{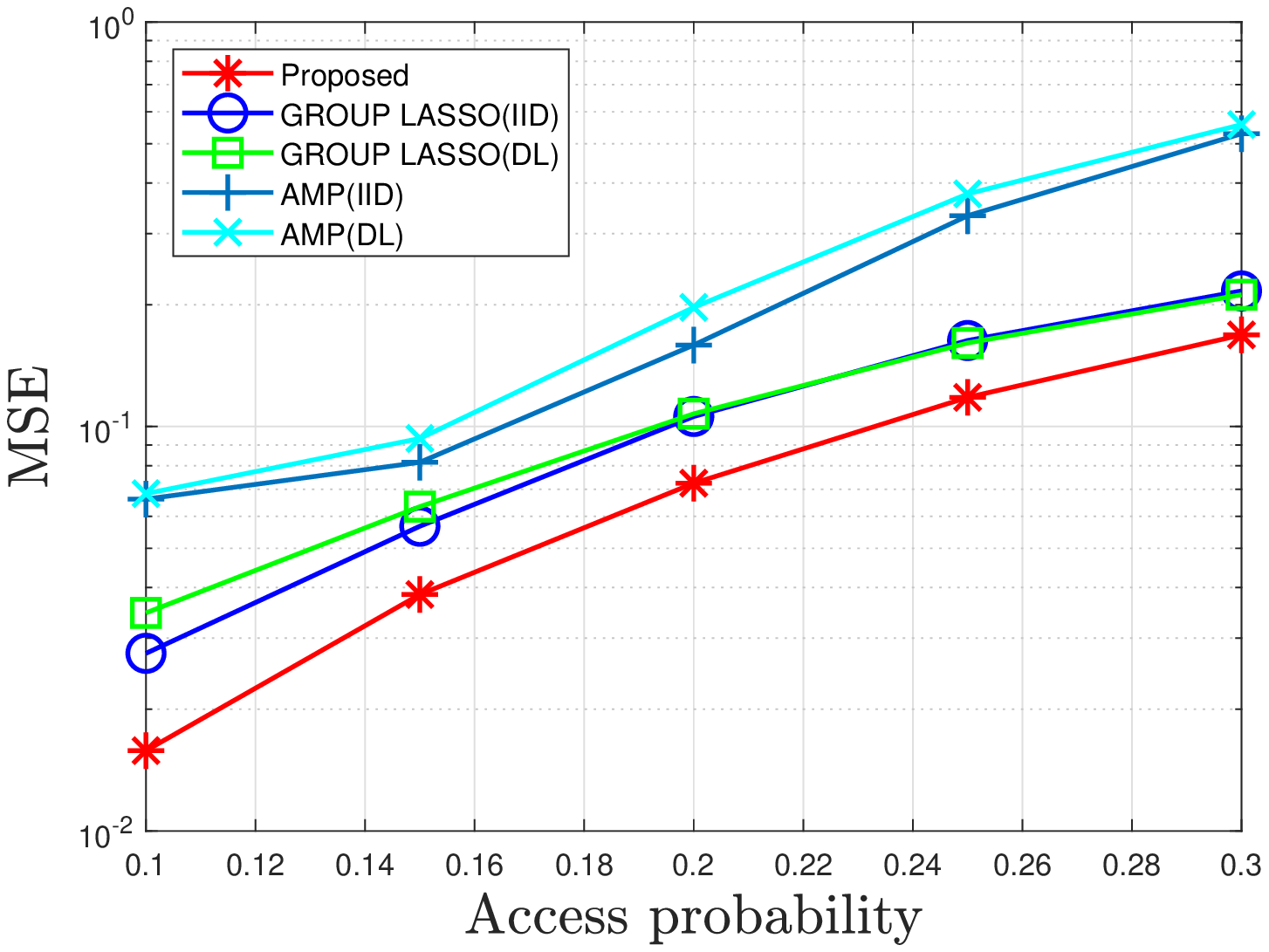}}}
 \subfigure[\scriptsize{MSE versus $L/N$ at $p=0.1$, $M=4$ in the correlated case.}]
 {\resizebox{4.2cm}{!}{\includegraphics{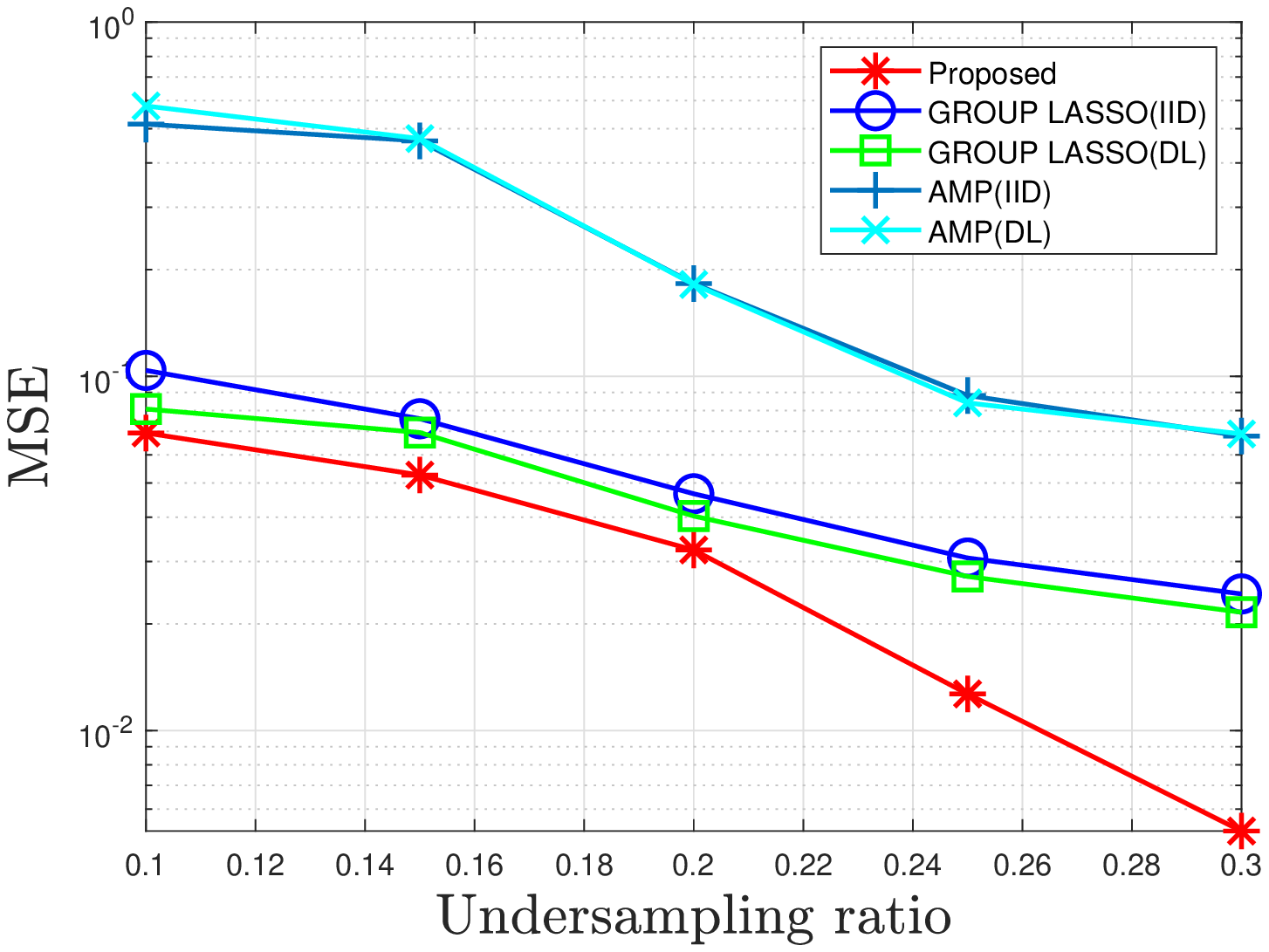}}}
 \subfigure[\scriptsize{MSE versus $p$ at $L/N=0.3$, $M=4$ in the correlated case.}]
 {\resizebox{4.2cm}{!}{\includegraphics{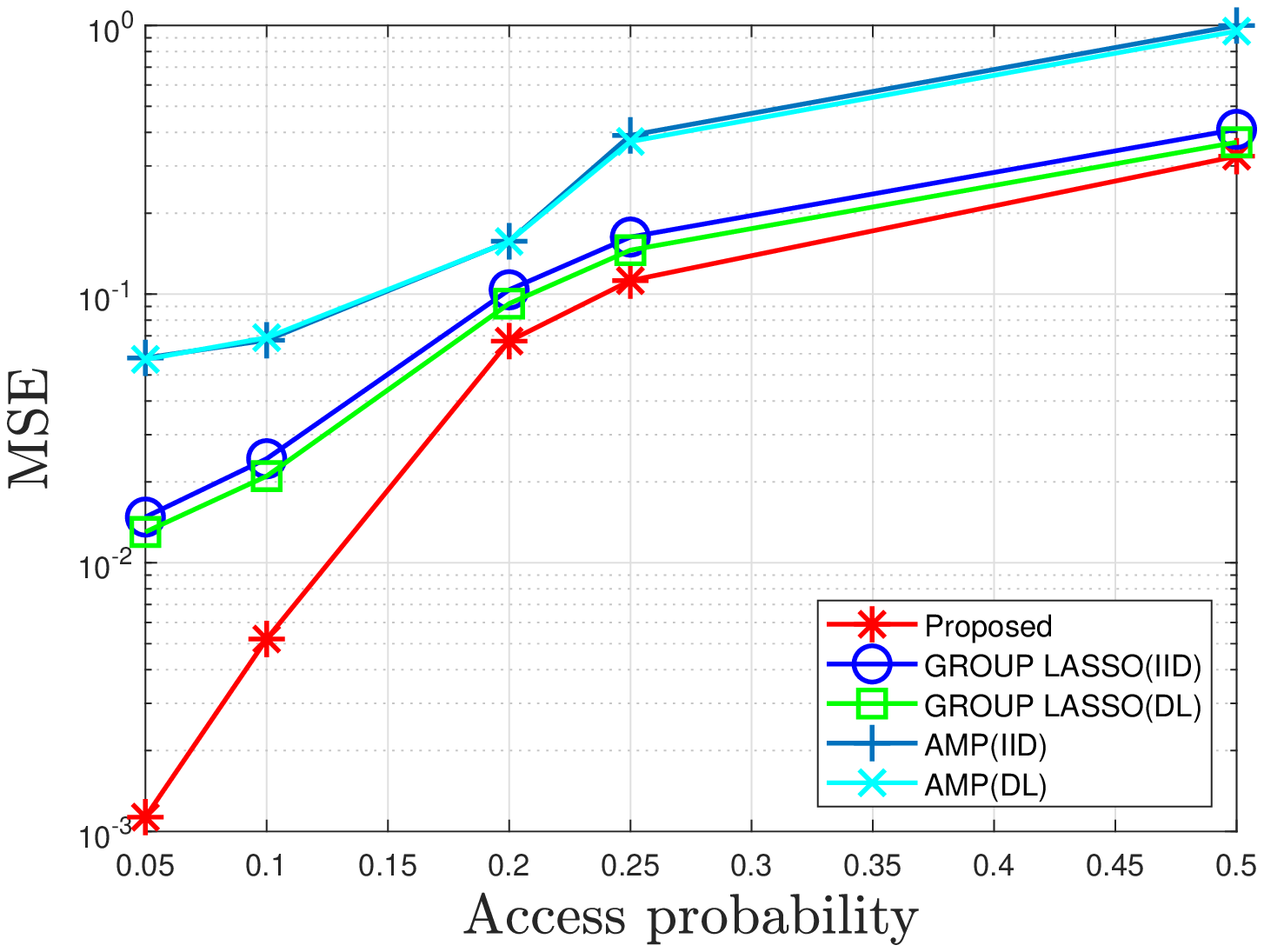}}}
  \end{center}
   \vspace{-0.3cm}
   \caption{\small{MSE versus undersampling ratio ($L/N$) and access probability ($p$) at $N=100$.}}
   \label{100}
   \vspace{-0.5cm}
\end{figure}

\begin{figure}[t]
\begin{center}
 \subfigure[\scriptsize{MSE versus $L/N$ at $p=0.1$, $M=4$ in the i.i.d. case.}]
 {\resizebox{4.2cm}{!}{\includegraphics{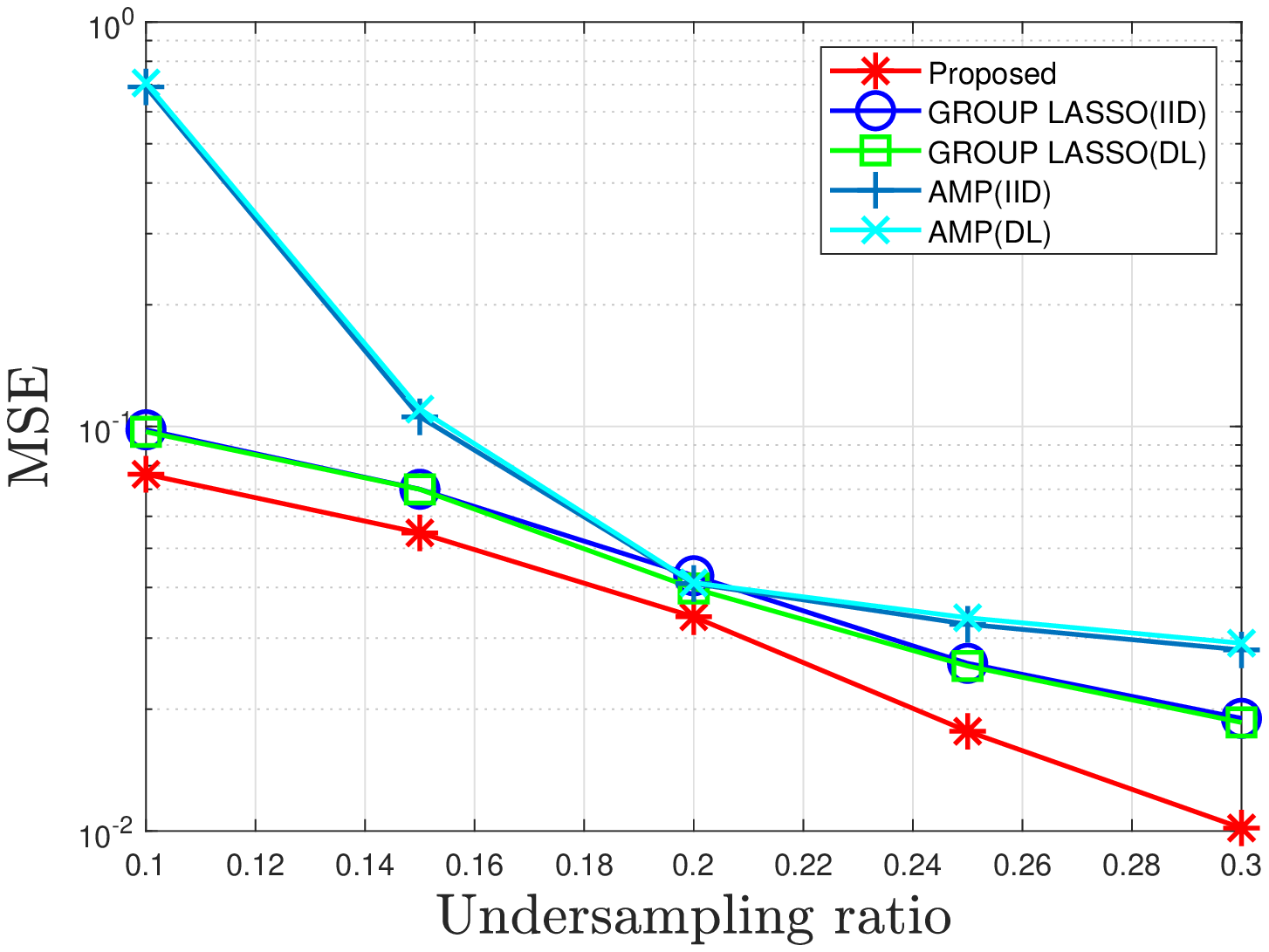}}}
 \subfigure[\scriptsize{MSE versus $p$ at $L/N=0.3$, $M=4$ in the i.i.d. case.}]
 {\resizebox{4.2cm}{!}{\includegraphics{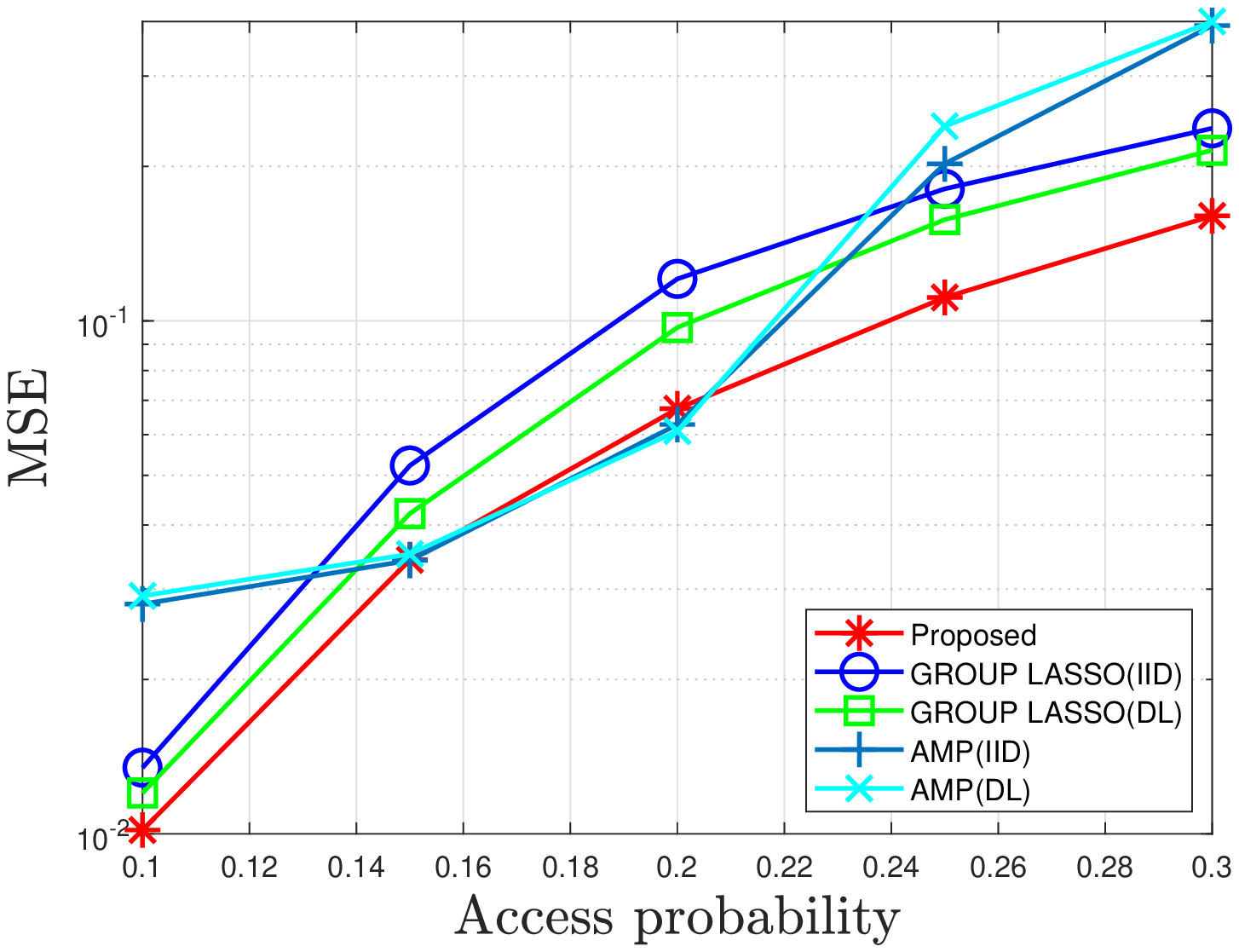}}}
 \subfigure[\scriptsize{MSE versus $L/N$ at $p=0.1$, $M=4$ in the correlated case.}]
 {\resizebox{4.2cm}{!}{\includegraphics{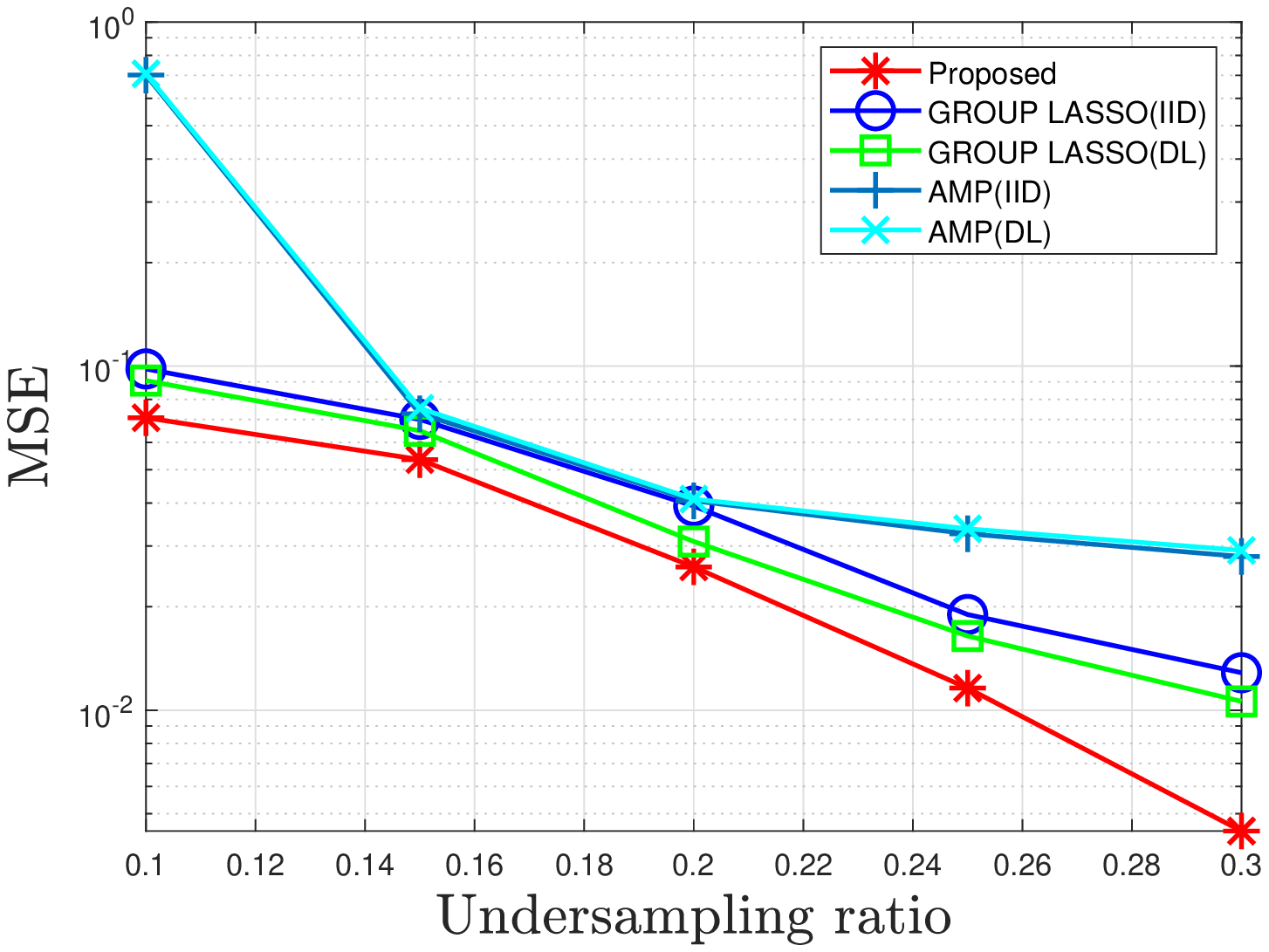}}}
 \subfigure[\scriptsize{MSE versus $p$ at $L/N=0.3$, $M=4$ in the correlated case.}]
 {\resizebox{4.2cm}{!}{\includegraphics{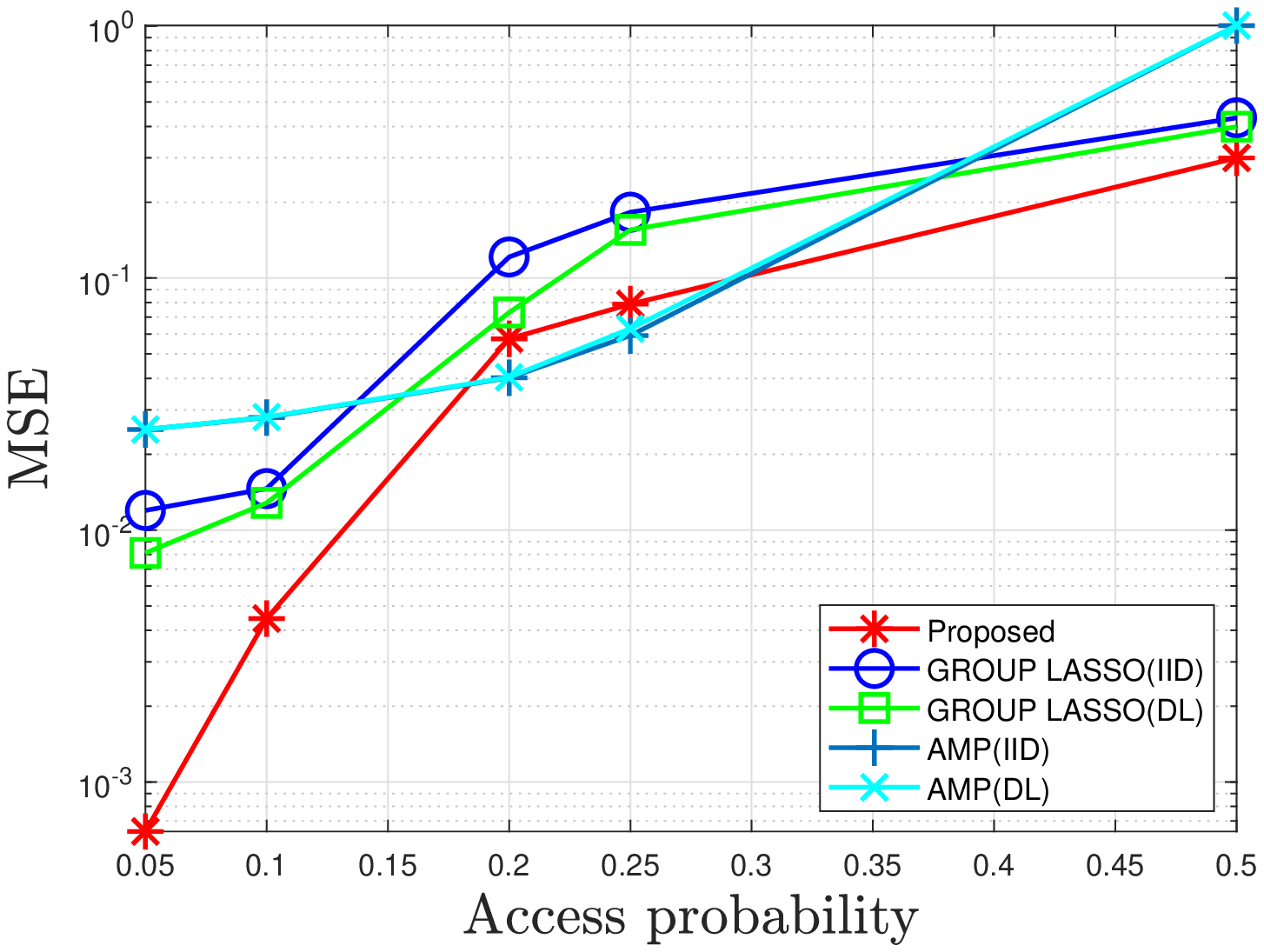}}}
  \end{center}
   \vspace{-0.5cm}
   \caption{\small{MSE versus undersampling ratio ($L/N$) and access probability ($p$) at $N=500$.}}
   \label{500}
\end{figure}

\begin{figure}[t]
\begin{center}
  \subfigure[\scriptsize{Time versus $L/N$ at $p=0.1$, $M=4$, $N=100$ in the i.i.d. case.}]
 {\resizebox{4.2cm}{!}{\includegraphics{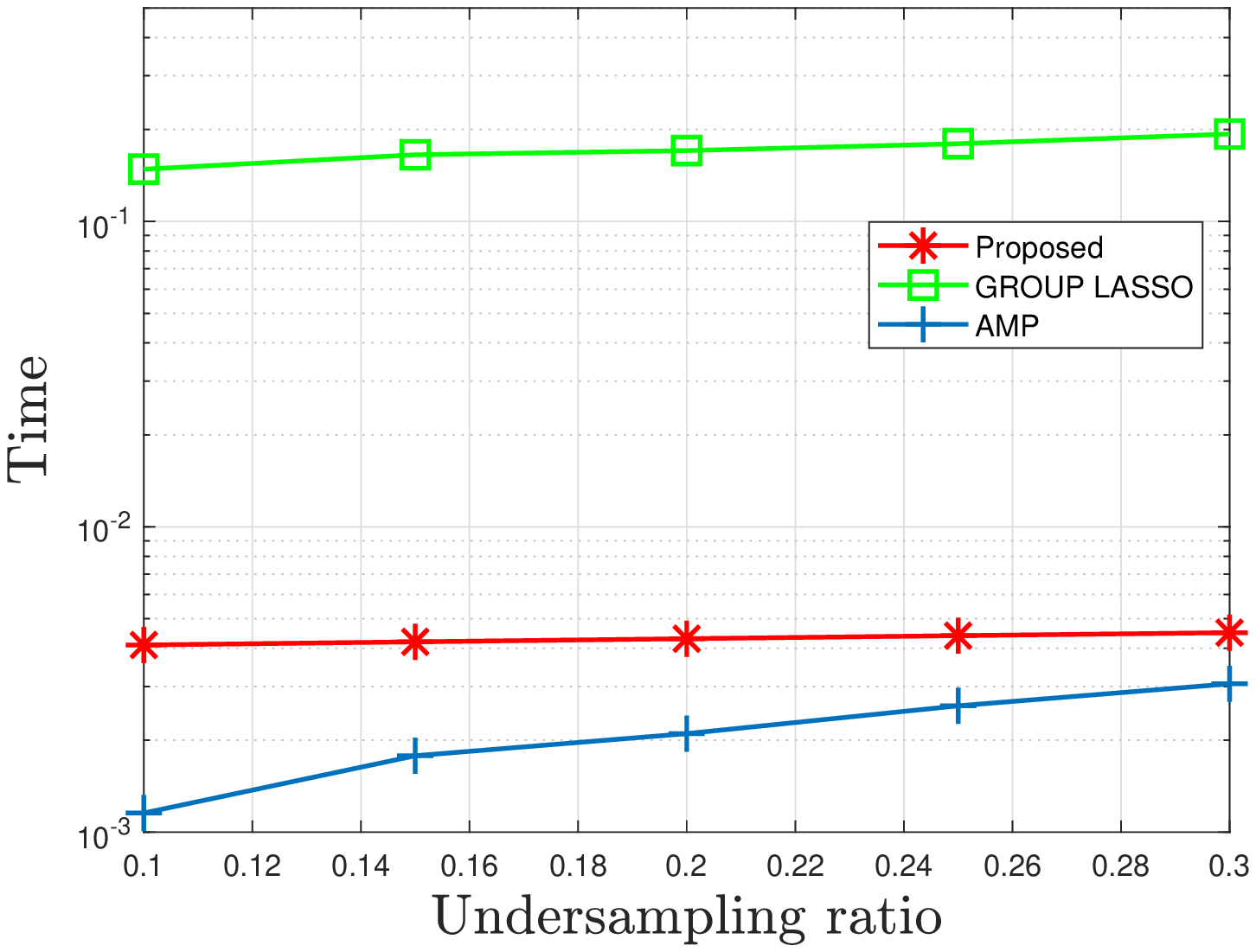}}}
  \subfigure[\scriptsize{Time versus $L/N$ at $p=0.1$, $M=4$, $N=500$ in the i.i.d. case.}]
 {\resizebox{4.2cm}{!}{\includegraphics{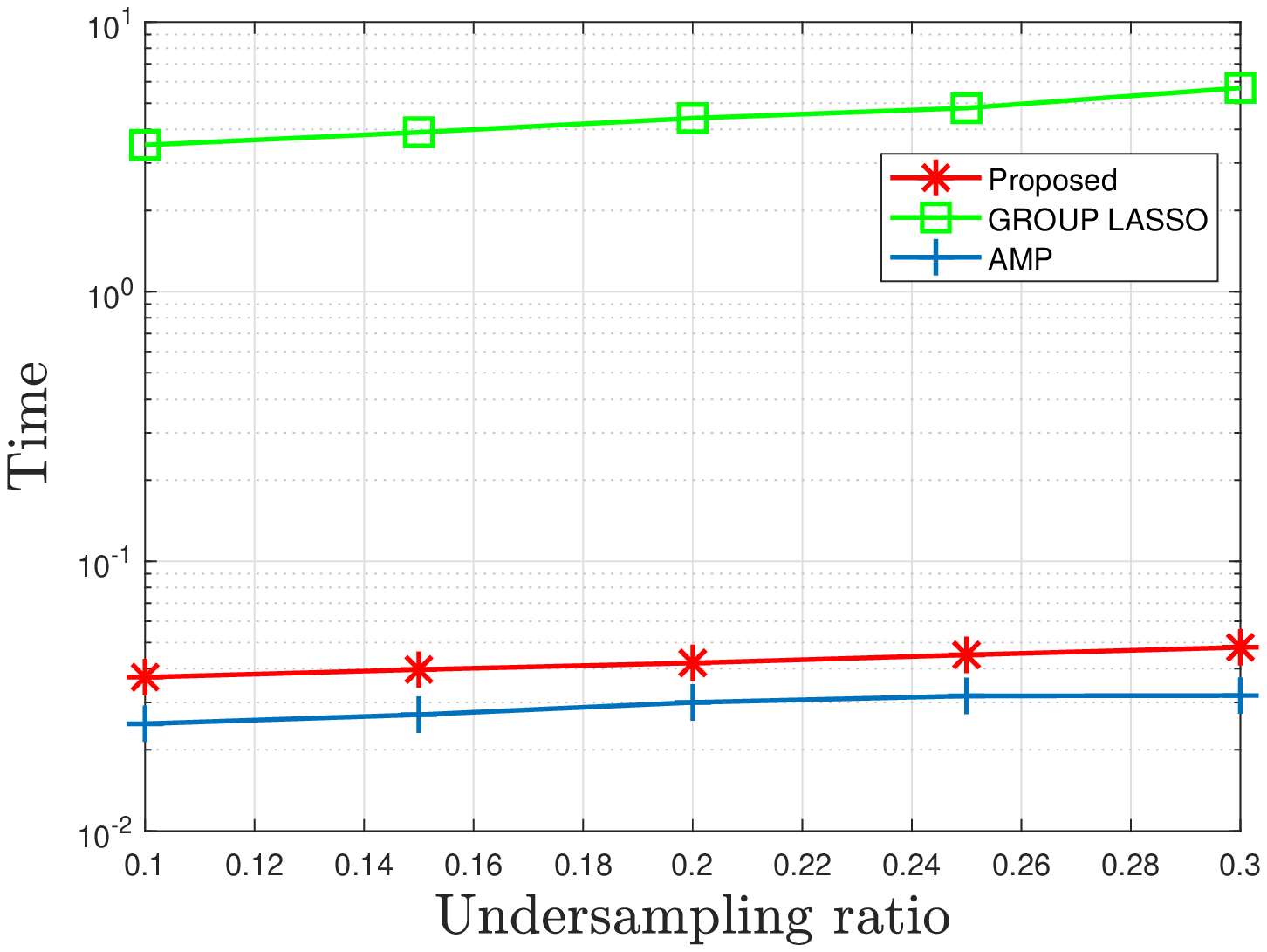}}}
  \end{center}
   \vspace{-0.5cm}
   \caption{\small{Computation time (in seconds) versus undersampling ratio ($L/N$).}}
   \label{time}
\end{figure}

Fig.~\ref{100} and Fig.~\ref{500} illustrate the MSE versus the undersampling ratio $L/N$ and access probability $p$ in both cases at $N=100$ and $N=500$, respectively. From Fig.~\ref{100} and Fig.~\ref{500}, we can see that in each case GROUP LASSO (DL) achieves a smaller MSE than GROUP LASSO (IID), which demonstrates that the pilot sequences obtained by the proposed model-driven approach are also effective for GROUP LASSO. The reason is as follows. The encoder and the GROUP LASSO-based decoder in the proposed architecture are jointly trained, and hence the obtained pilot sequences fit for GROUP LASSO to certain extent. In contrast, AMP (DL) has a larger MSE than AMP (IID), which demonstrates that the pilot sequences obtained by the proposed model-driven approach do not work for AMP. It is worth noting that the obtained pilot sequences are in general not suitable for unrelated sparse signal recovery methods. In addition, from Fig.~\ref{100} and Fig.~\ref{500}, we can see that GROUP LASSO achieves a much smaller MSE than AMP at all considered undersampling ratios and access probabilities when $N=100$; GROUP LASSO achieves a smaller MSE than AMP only at some considered undersampling ratios and access probabilities when $N=500$. This shows that GROUP LASSO is preferable when $N$ is not very large. It is known that AMP is especially suitable for the situation when $N$ is large.

Fig.~\ref{100} and Fig.~\ref{500} show that the proposed model-driven approach achieves a much smaller MSE than GROUP LASSO at all considered parameters, demonstrating the benefit of the correction layers in the GROUP LASSO-based decoder in reducing MSE. Similarly, the proposed model-driven approach achieves a much smaller MSE than AMP at all considered undersampling ratios and access probabilities when $N=100$, and achieves a smaller MSE than AMP at most of the considered parameters (except for $p=0.2$ in the i.i.d. case and $p=0.2, 0.25$ in the correlated case). This indicates that it is more suitable to use the proposed model-driven approach when $N$ is not too large. We can also see that the gains of the proposed model-driven approach over GROUP LASSO and AMP in the correlated case are larger than those in the i.i.d. case, which indicates that the proposed model-driven approach can exploit the properties of sparsity patterns for further reducing MSE.

Fig.~\ref{time} illustrates the computation time (in seconds) versus the undersampling ratio. From Fig.~\ref{time}, we can see that the computation time of the proposed model-driven approach is about 2.5 percent of that of GROUP LASSO, owning to higher computation efficiency of PCD-MMV; and the computation time of the proposed model-driven approach is only slightly longer than that of AMP. Note that the computation time of each scheme depends (almost) only on $N$ and $L$, and (almost) does not change with the sparsity pattern. In addition, it is worth noting that when $N$ is not very large, the MSE of AMP is unsatisfactory.

\section{Conclusion}
\label{sec:majhead}
Utilizing techniques in compressed sensing, parallel optimization, and deep learning, we propose a model-driven approach to jointly design the common measurement matrix and jointly sparse signal recovery method for complex signals, based on the standard auto-encoder structure for real numbers. The GROUP LASSO-based decoder consists of an approximation part which unfolds (several iterations of) the PCD-MMV algorithm to obtain an approximate solution of GROUP LASSO and a correction part which reduces the difference between the approximate solution and the actual jointly sparse signals. The proposed model-driven approach achieves higher recovery accuracy with less computation time than the GROUP LASSO method. In addition, the proposed model-driven approach can effectively utilize the properties of sparse patterns for improving recovery accuracy. This work highlights the value of parallel optimization in deep learning. Furthermore, the proposed model-driven approach can be extended to design common measurement matrices for other state-of-the-art jointly sparse signal recovery methods and reduce the computation times of those methods.


\begin{thebibliography}{10}
\providecommand{\url}[1]{#1}
\csname url@samestyle\endcsname
\providecommand{\newblock}{\relax}
\providecommand{\bibinfo}[2]{#2}
\providecommand{\BIBentrySTDinterwordspacing}{\spaceskip=0pt\relax}
\providecommand{\BIBentryALTinterwordstretchfactor}{4}
\providecommand{\BIBentryALTinterwordspacing}{\spaceskip=\fontdimen2\font plus
\BIBentryALTinterwordstretchfactor\fontdimen3\font minus
  \fontdimen4\font\relax}
\providecommand{\BIBforeignlanguage}[2]{{%
\expandafter\ifx\csname l@#1\endcsname\relax
\typeout{** WARNING: IEEEtran.bst: No hyphenation pattern has been}%
\typeout{** loaded for the language `#1'. Using the pattern for}%
\typeout{** the default language instead.}%
\else
\language=\csname l@#1\endcsname
\fi
#2}}
\providecommand{\BIBdecl}{\relax}
\BIBdecl

\bibitem{qin2013efficient}
Z.~Qin, K.~Scheinberg, and D.~Goldfarb, ``Efficient block-coordinate descent
  algorithms for the group lasso,'' \emph{Mathematical Programming
  Computation}, vol.~5, no.~2, pp. 143--169, 2013.

\bibitem{8323218}
L.~{Liu} and W.~{Yu}, ``Massive connectivity with massive {MIMO}-{Part I}:
  Device activity detection and channel estimation,'' \emph{IEEE Trans. Signal
  Process.}, vol.~66, no.~11, pp. 2933--2946, Jun 2018.

\bibitem{8264818}
Z.~{Chen}, F.~{Sohrabi}, and W.~{Yu}, ``Sparse activity detection for massive
  connectivity,'' \emph{IEEE Trans. Signal Process.}, vol.~66, no.~7, pp.
  1890--1904, April 2018.

\bibitem{senel2018grant}
K.~Senel and E.~G. Larsson, ``Grant-free massive mtc-enabled massive mimo: A
  compressive sensing approach,'' \emph{IEEE Trans. Commun.}, vol.~66, no.~12,
  pp. 6164--6175, 2018.

\bibitem{koochakzadeh2018fundamental}
A.~Koochakzadeh, H.~Qiao, and P.~Pal, ``On fundamental limits of joint sparse
  support recovery using certain correlation priors,'' \emph{IEEE Trans. Signal
  Process.}, vol.~66, no.~17, pp. 4612--4625, 2018.

\bibitem{8437359}
S.~{Haghighatshoar}, P.~{Jung}, and G.~{Caire}, ``Improved scaling law for
  activity detection in massive {MIMO} systems,'' in \emph{IEEE ISIT}, Jun
  2018, pp. 381--385.

\bibitem{obozinski2011support}
G.~Obozinski, M.~J. Wainwright, M.~I. Jordan \emph{et~al.}, ``Support union
  recovery in high-dimensional multivariate regression,'' \emph{The Annals of
  Statistics}, vol.~39, no.~1, pp. 1--47, 2011.

\bibitem{liu2018sparse}
L.~Liu, E.~G. Larsson, W.~Yu, P.~Popovski, C.~Stefanovic, and E.~De~Carvalho,
  ``Sparse signal processing for grant-free massive connectivity: A future
  paradigm for random access protocols in the internet of things,'' \emph{IEEE
  Signal Process. Mag.}, vol.~35, no.~5, pp. 88--99, 2018.

\bibitem{sun2016deep}
B.~Sun, H.~Feng, K.~Chen, and X.~Zhu, ``A deep learning framework of quantized
  compressed sensing for wireless neural recording,'' \emph{IEEE Access},
  vol.~4, pp. 5169--5178, 2016.

\bibitem{wu2019learning}
S.~Wu, A.~Dimakis, S.~Sanghavi, F.~Yu, D.~Holtmann-Rice, D.~Storcheus,
  A.~Rostamizadeh, and S.~Kumar, ``Learning a compressed sensing measurement
  matrix via gradient unrolling,'' in \emph{ICML}, 2019, pp. 6828--6839.

\bibitem{nguyen2017deep}
D.~M. Nguyen, E.~Tsiligianni, and N.~Deligiannis, ``Deep learning sparse
  ternary projections for compressed sensing of images,'' in \emph{IEEE
  GlobalSIP}, 2017, pp. 1125--1129.

\bibitem{8262812}
A.~{Mousavi}, G.~{Dasarathy}, and R.~G. {Baraniuk}, ``Deepcodec: Adaptive
  sensing and recovery via deep convolutional neural networks,'' in \emph{55th
  Allerton}, Oct 2017, pp. 744--744.

\bibitem{shi2017deep}
W.~Shi, F.~Jiang, S.~Zhang, and D.~Zhao, ``Deep networks for compressed image
  sensing,'' in \emph{IEEE ICME}, 2017, pp. 877--882.

\bibitem{adler2017block}
A.~Adler, D.~Boublil, and M.~Zibulevsky, ``Block-based compressed sensing of
  images via deep learning,'' in \emph{IEEE MMSP}, 2017, pp. 1--6.

\bibitem{8322184}
C.~{Wen}, W.~{Shih}, and S.~{Jin}, ``Deep learning for massive {MIMO} {CSI}
  feedback,'' \emph{IEEE Commun. Lett}, vol.~7, no.~5, pp. 748--751, Oct 2018.

\bibitem{8861085}
S.~{Li}, W.~{Zhang}, Y.~{Cui}, H.~{Cheng}, and W.~{Yu}, ``Joint design of
  measurement matrix and sparse support recovery method via deep
  auto-encoder,'' \emph{IEEE Signal Process. Lett}, pp. 1--1, Oct 2019.

\bibitem{7547360}
Y.~{Sun}, P.~{Babu}, and D.~P. {Palomar}, ``Majorization-minimization
  algorithms in signal processing, communications, and machine learning,''
  \emph{IEEE Transactions on Signal Processing}, vol.~65, no.~3, pp. 794--816,
  Feb 2017.

\end{thebibliography}


\end{document}